%mp_arc  cond-mat/
%**start of header
\newcount\mgnf  %ingrandimento
\mgnf=1

\ifnum\mgnf=0
\magnification=1000   
\hsize=15truecm\vsize=21truecm%\voffset2.truecm\hoffset.5truecm
   \parindent=0.3cm\baselineskip=0.45cm\fi
\ifnum\mgnf=1
   \magnification=\magstephalf
%\voffset=.5truecm
 %  \hoffset=0.truecm
 %  \hsize=15truecm\vsize=20.2truecm
   \baselineskip=18truept plus0.1pt minus0.1pt \parindent=0.9truecm
 %  \lineskip=0.5truecm\lineskiplimit=0.1pt      \parskip=0.1pt plus1pt
\fi

%\ifnum\mgnf=2%
%   \magnification=1200%
%   \hsize=15truecm\vsize=20.2truecm%
%   \baselineskip=18truept plus0.1pt minus0.1pt \parindent=0.9truecm%
%   \lineskip=0.5truecm\lineskiplimit=0.1pt      \parskip=0.1pt plus1pt%
%\fi
\ifnum\mgnf=2\magnification=1200\fi

\ifnum\mgnf=0
\def\openone{\leavevmode\hbox{\ninerm 1\kern-3.3pt\tenrm1}}%
\def\*{\vskip.7truemm}\fi
\ifnum\mgnf=1
\def\openone{\leavevmode\hbox{\ninerm 1\kern-3.63pt\tenrm1}}%
\def\*{\vglue0.3truecm}\fi
\ifnum\mgnf=2\def\*{\vglue0.7truecm}\fi
\openout15=\jobname.aux

%%%%%%%%%%%%%%%%%%%%%%%%%%%%%%%%%%%%%%%%%%%%%%%%%%%%%%%%%%%%%%%%%%%%%%%%%%%%%
%%%%%%%%%%%%%%%%%%%%%%%%%  DEFINIZIONI DI FONT    %%%%%%%%%%%%%%%%%%%%%%%%%%%
%%%%%%%%%%%%%%%%%%%%%%%%%%%%%%%%%%%%%%%%%%%%%%%%%%%%%%%%%%%%%%%%%%%%%%%%%%%%%
\font\titolo=cmbx12\font\titolone=cmbx10 scaled\magstep 2%
\font\sc=cmcsc10\font\css=cmcsc8%
\font\indbf=cmbx10 scaled\magstep2
\font\ottorm=cmr8\font\ninerm=cmr9%
\font\msytw=msbm9 scaled\magstep1%
%
%
%\font\msytwwww=msbm4 scaled\magstep1%
%
%
%
\def\st{\scriptstyle}%
%

%%%%%%%%%%%%%%%%%%%%%%%%%%%%%%%%%%%%%%%%%%%%%%%%%%%%%%%%%%%%%%%%%%%%%%%%%%%%%
%%%%%%%%%%%%%%%%%    LETTERE GRECHE E LATINE IN NERETTO     %%%%%%%%%%%%%%%%%
%%%%%%%%%%%%%%%%%%%%%%%%%%%%%%%%%%%%%%%%%%%%%%%%%%%%%%%%%%%%%%%%%%%%%%%%%%%%%

% lettere greche e latine in neretto italico - pag.430 del manuale
\font\tenmib=cmmib10 \font\eightmib=cmmib8
\font\sevenmib=cmmib7\font\fivemib=cmmib5 
\font\ottoit=cmti8
\font\fiveit=cmti5\font\sixit=cmti6%%
%!!!@@@\font\fiveit=cmti7\font\sixit=cmti7%%
\font\fivei=cmmi5\font\sixi=cmmi6\font\ottoi=cmmi8
\font\ottorm=cmr8
\font\ottosy=cmsy8\font\sixsy=cmsy6\font\fivesy=cmsy5%%
\font\ottobf=cmbx8\font\sixbf=cmbx6\font\fivebf=cmbx5%
\font\ottocss=cmcsc8%

\def\ottopunti{\def\rm{\fam0\ottorm}\def\it{\fam6\ottoit}%
\def\bf{\fam7\ottobf}%
\textfont1=\ottoi\scriptfont1=\sixi\scriptscriptfont1=\fivei%
\textfont2=\ottosy\scriptfont2=\sixsy\scriptscriptfont2=\fivesy%
%\textfont3=\tenex\scriptfont3=\tenex\scriptscriptfont3=\tenex%
\textfont4=\ottocss\scriptfont4=\sc\scriptscriptfont4=\sc%
%\scriptfont4=\ottocss\scriptscriptfont4=\ottocss%
\textfont5=\eightmib\scriptfont5=\sevenmib\scriptscriptfont5=\fivemib%
\textfont6=\ottoit\scriptfont6=\sixit\scriptscriptfont6=\fiveit%
\textfont7=\ottobf\scriptfont7=\sixbf\scriptscriptfont7=\fivebf%
%\textfont\bffam=\eightmib\scriptfont\bffam=\sevenmib%
%\scriptscriptfont\bffam=\fivemib%
\setbox\strutbox=\hbox{\vrule height7pt depth2pt width0pt}%
\normalbaselineskip=9pt\rm}

\textfont5=\tenmib\scriptfont5=\sevenmib\scriptscriptfont5=\fivemib
\mathchardef\Ba   = "050B  %alfa
\mathchardef\Bb   = "050C  %beta
\mathchardef\Bg   = "050D  %gamma
\mathchardef\Bd   = "050E  %delta
\mathchardef\Be   = "0522  %varepsilon
\mathchardef\Bee  = "050F  %epsilon
\mathchardef\Bz   = "0510  %zeta
\mathchardef\Bh   = "0511  %eta
\mathchardef\Bthh = "0512  %teta
\mathchardef\Bth  = "0523  %varteta
\mathchardef\Bi   = "0513  %iota
\mathchardef\Bk   = "0514  %kappa
\mathchardef\Bl   = "0515  %lambda
\mathchardef\Bm   = "0516  %mu
\mathchardef\Bn   = "0517  %nu
\mathchardef\Bx   = "0518  %xi
\mathchardef\Bom  = "0530  %omi
\mathchardef\Bp   = "0519  %pi
\mathchardef\Br   = "0525  %ro
\mathchardef\Bro  = "051A  %varrho
\mathchardef\Bs   = "051B  %sigma
\mathchardef\Bsi  = "0526  %varsigma
\mathchardef\Bt   = "051C  %tau
\mathchardef\Bu   = "051D  %upsilon
\mathchardef\Bf   = "0527  %phi
\mathchardef\Bff  = "051E  %varphi
\mathchardef\Bch  = "051F  %chi
\mathchardef\Bps  = "0520  %psi
\mathchardef\Bo   = "0521  %omega
\mathchardef\Bome = "0524  %varomega
\mathchardef\BG   = "0500  %Gamma
\mathchardef\BD   = "0501  %Delta
\mathchardef\BTh  = "0502  %Theta
\mathchardef\BL   = "0503  %Lambda
\mathchardef\BX   = "0504  %Xi
\mathchardef\BP   = "0505  %Pi
\mathchardef\BS   = "0506  %Sigma
\mathchardef\BU   = "0507  %Upsilon
\mathchardef\BF   = "0508  %Fi
\mathchardef\BPs  = "0509  %Psi
\mathchardef\BO   = "050A  %Omega
\mathchardef\BDpr = "0540  %Dpr
\mathchardef\Bstl = "053F  %*

%%%%%%%%%%%%%%%%%%%%%%%%%%%%%%%%%%%%%%%%%%%%%%%%%%%%%%%%%%%%%%%%%%%%%%%%%%%%%
%%%%%%%%%   RIFERIMENTI SIMBOLICI A FORMULE, PARAGRAFI E FIGURE    %%%%%%%%%%
%%%%%%%%%%%%%%%%%%%%%%%%%%%%%%%%%%%%%%%%%%%%%%%%%%%%%%%%%%%%%%%%%%%%%%%%%%%%%
%
% Ogni paragrafo deve iniziare con il comando \section(#1,#2), dove #1
% e' il simbolo associato al paragrafo e #2 e' il titolo. Per le
% appendici bisogna pero' usare \appendix(#1,#2).
%
% Se nel titolo compaiono riferimenti ad altri simboli, questi vanno
% racchiusi fra parentesi graffe, per es. {\equ(1.2)}; in caso contrario
% si provoca un errore.
%
% Ogni sottoparagrafo deve iniziare con il comando \sub(#1) o \asub(#1),
% nelle appendici.
%
% I riferimenti a paragrafi e sottoparagrafi si realizzano con il comando
% \sec(#1), che produce il numero effettivo preceduto dal simbolo di
% paragrafo, o \secc(#1), che produce solo il numero (serve nel caso si
% faccia riferimento ad un sottoparagrafo, che e' un Lemma, un Teorema o
% altro oggetto suscettibile di una denominazione speciale).
%
% Le formule sono contrassegnate con \Eq(#1), eccetto che all'interno
% del comando \eqalignno, dove si deve usare \eq(#1). Nelle appendici
% i comandi corrispondenti sono \Eqa(#1) e \eqa(#1).
% I riferimenti alle formule si realizzano con \equ(#1).
%
% La numerazione delle figure utilizza il comando \eqg(#1), per
% contrassegnarle, e \graf(#1) per citarle.
%

\global\newcount\numsec\global\newcount\numapp
\global\newcount\numfor\global\newcount\numfig
\global\newcount\numsub
\numsec=0\numapp=0\numfig=0
\def\veroparagrafo{\number\numsec}\def\veraformula{\number\numfor}
\def\veraappendice{\number\numapp}\def\verasub{\number\numsub}
\def\verafigura{\number\numfig}

\def\Section(#1,#2){\advance\numsec by 1\numfor=1\numsub=1\numfig=1%
\SIA p,#1,{\veroparagrafo} %
\write15{\string\Fp (#1){\secc(#1)}}%
\write16{ sec. #1 ==> \secc(#1)  }%
\0\hbox%to \hsize
{\titolo\hfill
\number\numsec. #2\hfill%
\expandafter{\hglue-1truecm\alato(sec. #1)}}}

\def\appendix(#1,#2){\advance\numapp by 1\numfor=1\numsub=1\numfig=1%
\SIA p,#1,{A\veraappendice} %
\write15{\string\Fp (#1){\secc(#1)}}%
\write16{ app. #1 ==> \secc(#1)  }%
\hbox to \hsize{\titolo Appendix A\number\numapp. #2\hfill%
\expandafter{\alato(app. #1)}}\*%
}

\def\senondefinito#1{\expandafter\ifx\csname#1\endcsname\relax}

\def\SIA #1,#2,#3 {\senondefinito{#1#2}%
\expandafter\xdef\csname #1#2\endcsname{#3}\else
\write16{???? ma #1#2 e' gia' stato definito !!!!} \fi}

\def \Fe(#1)#2{\SIA fe,#1,#2 }
\def \Fp(#1)#2{\SIA fp,#1,#2 }
\def \Fg(#1)#2{\SIA fg,#1,#2 }

\def\etichetta(#1){(\veroparagrafo.\veraformula)%
\SIA e,#1,(\veroparagrafo.\veraformula) %
\global\advance\numfor by 1%
\write15{\string\Fe (#1){\equ(#1)}}%
\write16{ EQ #1 ==> \equ(#1)  }}

\def\etichettaa(#1){(A\veraappendice.\veraformula)%
\SIA e,#1,(A\veraappendice.\veraformula) %
\global\advance\numfor by 1%
\write15{\string\Fe (#1){\equ(#1)}}%
\write16{ EQ #1 ==> \equ(#1) }}

\def\getichetta(#1){\veroparagrafo.\verafigura%
\SIA g,#1,{\veroparagrafo.\verafigura} %
\global\advance\numfig by 1%
\write15{\string\Fg (#1){\graf(#1)}}%
\write16{ Fig. #1 ==> \graf(#1) }}

\def\etichettap(#1){\veroparagrafo.\verasub%
\SIA p,#1,{\veroparagrafo.\verasub} %
\global\advance\numsub by 1%
\write15{\string\Fp (#1){\secc(#1)}}%
\write16{ par #1 ==> \secc(#1)  }}

\def\Eq(#1){\eqno{\etichetta(#1)\alato(#1)}}
\def\eq(#1){\etichetta(#1)\alato(#1)}
\def\Eqa(#1){\eqno{\etichettaa(#1)\alato(#1)}}
\def\eqa(#1){\etichettaa(#1)\alato(#1)}
\def\eqg(#1){\getichetta(#1)\alato(fig. #1)}
\def\sub(#1){\0\palato(p. #1){\bf \etichettap(#1).}}
\def\asub(#1){\0\palato(p. #1){\bf \etichettapa(#1).}}
\def\apprif(#1){\senondefinito{e#1}%
\eqv(#1)\else\csname e#1\endcsname\fi}

\def\equv(#1){\senondefinito{fe#1}$\clubsuit$#1%
\write16{eq. #1 non e' (ancora) definita}%
\else\csname fe#1\endcsname\fi}
\def\grafv(#1){\senondefinito{fg#1}$\clubsuit$#1%
\write16{fig. #1 non e' (ancora) definito}%
\else\csname fg#1\endcsname\fi}
\def\secv(#1){\senondefinito{fp#1}$\clubsuit$#1%
\write16{par. #1 non e' (ancora) definito}%
\else\csname fp#1\endcsname\fi}

\def\eqo{{\global\advance\numfor by 1}}
\def\equ(#1){\senondefinito{e#1}\equv(#1)\else\csname e#1\endcsname\fi}
\def\graf(#1){\senondefinito{g#1}\grafv(#1)\else\csname g#1\endcsname\fi}
\def\figura(#1){{\css Figura} \getichetta(#1)}
\def\secc(#1){\senondefinito{p#1}\secv(#1)\else\csname p#1\endcsname\fi}
\def\sec(#1){{\secc(#1)}}
\def\refe(#1){{[\secc(#1)]}}

\def\BOZZA{%\bz=1
\def\alato(##1){\rlap{\kern-\hsize\kern-.5truecm{$\scriptstyle##1$}}}
\def\palato(##1){\rlap{\kern-.5truecm{$\scriptstyle##1$}}}
}

\def\alato(#1){}
\def\galato(#1){}
\def\palato(#1){}

%%%%%%%%%%%%%%%%%%%%%%%%%%%%%%%%%%%%%%%%%%%%%%%%%%%%%%%%%%%%%%%%%%%%%%%%%%%%%
%%%%%%%%%%%%%%%%%%%%      DATA E PIE' DI PAGINA        %%%%%%%%%%%%%%%%%%%%%%
%%%%%%%%%%%%%%%%%%%%%%%%%%%%%%%%%%%%%%%%%%%%%%%%%%%%%%%%%%%%%%%%%%%%%%%%%%%%%

{\count255=\time\divide\count255 by 60 \xdef\hourmin{\number\count255}
        \multiply\count255 by-60\advance\count255 by\time
   \xdef\hourmin{\hourmin:\ifnum\count255<10 0\fi\the\count255}}

\def\oramin{\hourmin }

\def\data{\number\day/\ifcase\month\or gennaio \or febbraio \or marzo \or
aprile \or maggio \or giugno \or luglio \or agosto \or settembre
\or ottobre \or novembre \or dicembre \fi/\number\year;\ \oramin}
\setbox200\hbox{$\scriptscriptstyle \data $}

%%%%%%%%%%%%%%%%%%%%%%%%%%%%%%%%%%%%%%%%%%%%%%%%%%%%%%%%%%%%%%%%%%%%%%%%%%%%%
%%%%%%%%%%%%%%%      INSERIMENTO FIGURE ( se si usa DVIPS )    %%%%%%%%%%%%%%
%%%%%%%%%%%%%%%%%%%%%%%%%%%%%%%%%%%%%%%%%%%%%%%%%%%%%%%%%%%%%%%%%%%%%%%%%%%%%
\newdimen\xshift \newdimen\xwidth \newdimen\yshift \newdimen\ywidth

\def\ins#1#2#3{\vbox to0pt{\kern-#2\hbox{\kern#1 #3}\vss}\nointerlineskip}

\def\eqfig#1#2#3#4#5{
\par\xwidth=#1 \xshift=\hsize \advance\xshift
by-\xwidth \divide\xshift by 2
\yshift=#2 \divide\yshift by 2%
%\line
{\hglue\xshift \vbox to #2{\vfil
#3 \includegraphics{#4.ps}
}\hfill\raise\yshift\hbox{#5}}}

\def\8{\write12}

\openin13=#1.aux \ifeof13 \relax \else
\input #1.aux \closein13\fi
\openin14=\jobname.aux \ifeof14 \relax \else
\input \jobname.aux \closein14 \fi
\immediate\openout15=\jobname.aux

%%%%%%%%%%%%%%%%%%%%%%%%%%%%%%%%%%%%%%%%%%%%%%%%%%%%%%%%%%%%%%%%%%%%%%%%%%%%%
%%%%%%%%%%%%%%%%%%%%%%         SIMBOLI VARI           %%%%%%%%%%%%%%%%%%%%%%%
%%%%%%%%%%%%%%%%%%%%%%%%%%%%%%%%%%%%%%%%%%%%%%%%%%%%%%%%%%%%%%%%%%%%%%%%%%%%%

\let\a=\alpha \let\b=\beta  \let\g=\gamma  \let\d=\delta 
\let\z=\zeta     \let\th=\theta \let\k=\kappa \let\l=\lambda
\let\m=\mu    \let\n=\nu             \let\r=\rho
\let\s=\sigma \let\t=\tau   \let\f=\varphi 
  \let\ps=\psi   
   \let\L=\Lambda 
    \let\Si=\Sigma \let\F=\Phi

\def\\{\hfill\break} \let\==\equiv

\let\io=\infty 
\def\Dpr{\BDpr\,}%\def\Dpr{\V\dpr\,}
\def\ap{{\it a priori\ }}
\let\0=\noindent

\def\media#1{{\langle#1\rangle}}
\def\ie{\hbox{\it i.e.\ }}\def\eg{\hbox{\it e.g.\ }}
\let\dpr=\partial

\def\tende#1{\,\vtop{\ialign{##\crcr\rightarrowfill\crcr
 \noalign{\kern-1pt\nointerlineskip} \hskip3.pt${\scriptstyle
 #1}$\hskip3.pt\crcr}}\,}
\def\circage{\lower2pt\hbox{$\,\buildrel > \over {\scriptstyle \sim}\,$}}
\def\otto{\,{\kern-1.truept\leftarrow\kern-5.truept\to\kern-1.truept}\,}
\def\fra#1#2{{#1\over#2}}

\def\EE{{\cal E}}  
  
  \def\II{{\cal I}}
\def\RR{{\cal R}}  
\def\AA{{\cal A}}

\def\T#1{{#1_{\kern-3pt\lower7pt\hbox{$\widetilde{}$}}\kern3pt}}
\def\VVV#1{{\underline #1}_{\kern-3pt
\lower7pt\hbox{$\widetilde{}$}}\kern3pt\,}
\def\W#1{#1_{\kern-3pt\lower7.5pt\hbox{$\widetilde{}$}}\kern2pt\,}

\def\etc{{\it etc}} 
  
\def\indica{\leaders \hbox to 0.5cm{\hss.\hss}\hfill}
\def\guida{\leaders\hbox to 1em{\hss.\hss}\hfill}

\def\qed{\raise1pt\hbox{\vrule height5pt width5pt depth0pt}}

\def\indic{\hbox{\raise-2pt \hbox{\indbf 1}}}

\def\RRR{\hbox{\msytw R}}

\def\defi{\,{\buildrel def\over=}\,}

\def\rhs{{\it r.h.s.}\ }

\def\sqr#1#2{{\vcenter{\vbox{\hrule height.#2pt%
        \hbox{\vrule width.#2pt height#1pt \kern#1pt%
          \vrule width.#2pt}%
        \hrule height.#2pt}}}}

\def\ig{\int}

\footline={\rlap{\hbox{\copy200}}\tenrm\hss \number\pageno\hss}
\def\V#1{{\bf#1}}

\def\fiat{}

%BOZZA
\def\asint#1{\,\vtop{\ialign{##\crcr\hss $\simeq$\hss\crcr
 \noalign{\kern-1pt\nointerlineskip} \hskip3.pt${\scriptstyle
 #1}$\hskip3.pt\crcr}}\,}

\fiat
%**end of header

%\vglue1.truecm
\centerline{\titolone Stationary nonequilibrium statistical mechanics}
\*

\centerline
{\it Giovanni Gallavotti}

\centerline
{\it I.N.F.N. Roma 1, Fisica Roma1}

\*
\Section(1, Nonequilibrium.)
\*

Systems in stationary nonequilibrium are mechanical systems subject to
nonconservative external forces and to thermostat forces which forbid
indefinite increase of the energy and allow reaching statistically
stationary states. A system $\Si$ is described by the positions and
velocities of its $n$ particles $\V X,\dot{\V X}$, with the particles
positions confined to a finite volume container $C_0$.

If $\V X=(\V x_1,\ldots$, $\V x_n)$ are the particles positions in a
Cartesian inertial system of coordinates, the equations of motion are
determined by their masses $m_i>0,\, i=1,\ldots,n$, by the
{\it potential energy} of interaction $V(\V x_1,\ldots,\V x_n)\=V(\V
X)$, by the external nonconservative forces $\V F_i(\V X,\BF)$ and by the {\it
thermostat} forces $-\Bth_i$ as

$$m_i {\ddot{\V x}}_i=-\dpr_{\V x_i} V(\V X)+ \V F_i(\V X; \BF)
-\Bth_i,\qquad i=1,\ldots,n\Eq(1.1)$$
where $\BF=(\f_1,\ldots,\f_q)$ are strength parameters on which the
external forces depend. All forces and potentials will be supposed
smooth, \ie analytic, in their variables aside from {\it possible}
impulsive elastic forces describing shocks, and with the property: $\V
F(\V X;\V 0)=\V0$. The impulsive forces are allowed here to model
possible shocks with the walls of the container $C_0$ or between hard
core particles.

A kind of thermostats are {\it reservoirs} which may consist of one or
more {\it infinite} systems which are asymptotically in thermal
equilibrium and are separated by boundary surfaces from each other aas
well as from the system: with the latter they
interact through short range conservative forces, see Fig.1.
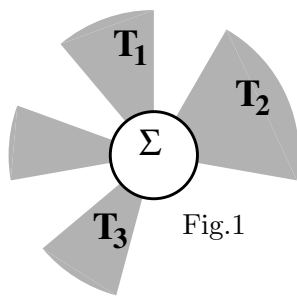
\eqfig{380pt}{100pt}
{\ins{110pt}{100pt}{\vbox{\hsize=250pt%
\0The reservoirs occupy infinite regions of the space outside $C_0$,
\eg sectors $C_a\subset \RRR^3$, $a=1,2\ldots$, in space and their
particles are in a configuration which is typical of an equilibrium
state at temperature $T_a$. This means that the {\it empirical}
probability of configurations in each $C_a$ is Gibbsian with some
temperature $T_a$. In other words the frequency with which a
configuration $(\dot {\V Y},\V Y+\V r)$ occurs in}}
\ins{60pt}{30pt}{Fig.1}
}{fig1}{}

\vskip4pt
\0a region $\L+\V
r\subset C_a$ and a configuration $(\dot{\V W},\V W+\V r)$ occurs
outside $\L+\V r$ (with $\V Y\subset \L,\V W\cap\L$\\$=\emptyset$)
{\it averaged over the translations $\L+\V r$ of $\L$ by $\V r$} (with
the restriction that $\L+\V r\subset C_a$) is

$${\rm average}\kern-30pt\lower8pt\hbox{$\st \V r+\L\subset C_a$}
(f_{\L+\V r}[(\dot{\V Y},\V Y+\V r); \dot{\V W},\V W+\V r])=
\fra{e^{-{\b_a}\big(\fra1{2m_a} | \dot{\V Y}|^2+ V_a(\V Y|\V
W)\big)}}{\rm normalization}\Eq(1.2)$$
here $m_a$ is the mass of the particles in the $a$-th reservoir and
$V_a(\V Y|\V W)$ is the energy of the short range potential between
pairs of particles in $\V Y\subset C_a$ or with one point in $\V Y$
and one in $\V W$. Since the configurations in the system and in the
thermostats are not random the \equ(1.2) should be considered as an
``empirical'' probability in the sense that it is the frequency
density of the events $\{(\dot{\V Y},\V Y+\V r); \V W+\V r\}$: in
other words the configurations $\Bo_a$ in the reservoirs should be
``typical'' in the sense of probability theory of distributions which
are asymptotically Gibbsian.

The property of being ``thermostats'' means that
\equ(1.2) remains true for all times, if initially satisfied.

Mathematically there is a problem at this point: the latter property
is either true or false, but a proof of its validity seems out of
reach of the present techniques except in very simple cases. Therefore
here we follow an intuitive approach and assume that such thermostats
exist and, actually, that any configuration which is typical of a
stationary state of an infinite size system of interacting particles
in the $C_a$'s, with physically reasonable microscopic interactions,
satisfies the property \equ(1.2).

The above thermostats are examples of ``deterministic thermostats''
because, together with the system they form a deterministic dynamical
system.  They are called ``Hamiltonian thermostats'' and are often
considered as the most appropriate models of ``physical thermostats''.

A closely related thermostat model is obtained by assuming that the
particles outside the system are not in a given configuration but they
have a probability distribution whose conditional distributions
satisfy \equ(1.2) initially. Also in this case it is necessary to
assume that \equ(1.2) remains true for all times, if initially
satisfied.  Such thermostats are examples of ``stochastic
thermostats'' because their action on the system depends on random
variables $\Bo_a$ which are the {\it initial} configurations of the
particles belonging to the thermostats.

Other kinds of stochastic thermostats are {\it collision rules} with
the container boundary $\dpr C_0$ of $\Si$: every time a
particle collides with $\dpr C_0$  it is reflected with a momentum
$\V p$ in $d^3\V p$ that has a probability distribution proportional
to $e^{-\b_a\fra1{2m}\V p^2}d^3\V p$ where $\b_a$, $a=1,2,\ldots$
depend on which boundary portion (labeled by $a=1,2\ldots$ and, if
$k_B$ is Boltzmann's constant, at ``temperature'' $T_a=(k_B
\b_a)^{-1}$) the collision takes place. Which $\V p$ is actually
chosen after each collision is determined by a random variable
$\Bo=(\Bo_1,\Bo_2,\ldots)$.

It is also possible, and convenient, to consider deterministic
thermostats which are {\it finite}. In the latter case
$\Bth$ is a force {\it only} depending upon the configuration of the
$n$ particles in their finite container $C_0$. The distinction between
stochastic and deterministic thermostats ultimately rests on what we
call ``system''. If reservoirs or the randomness generators are
included in the system then the system becomes deterministic (possibly
infinite); and finite deterministic thermostats can also regarded as
simplified models for infinite reservoirs, see Sect.\sec(4).

Examples of finite deterministic reservoirs are forces obtained by
imposing a nonholonomic constraint via some {\it ad hoc} principle
like the Gauss' principle. For instance if a system of particles
driven by a force $\V G_i\defi-\Dpr_{\V x_i}V(\V X)+\V F_i(\V X)$ is
enclosed in a box $C_0$ and $\Bth$ is a thermostat enforcing an
anholonomic constraint $\ps(\dot{\V X},\V X)\=0$ via Gauss' principle
then

$$\Bth_i(\dot{\V X},\V X)=
\Big[\fra{\sum_j \dot{\V x}_j\cdot\Dpr_{\V
x_j}\ps(\dot{\V X},\V X)+
\fra1m \V G_j\cdot \Dpr_{\dot{\V x}_j}\ps(\dot{\V X},\V X)} 
{\sum_j \fra1m (\Dpr_{\dot{\V
x}_j}\ps(\dot{\V X},\V X))^2}\Big]\, \Dpr_{\dot{\V
x}_i}\ps(\dot{\V X},\V X)
\Eq(1.3)$$
Gauss' principle is remarkable as it says that the force which needs
to be added to the other forces $\V G_i$ acting on the system minimizes $\sum_i
\fra{(\V G_i- m_i\V a_i)^2}{m_i}$, given $\dot{\V X},\V X$,
among all accelerations $\V a_i$ which are compatible with the
constraint $\ps$.

{\it For simplicity stochastic or infinite thermostats will not be
considered} here. It should be kept in mind that the only known
examples of mathematically treatable thermostats modeled by infinite
reservoirs are cases in which the thermostats particles are either
noninteracting particles or linear (\ie noninteracting) oscillators.

In general in order that a force $\Bth$ can be considered a
deterministic ``thermostat force'' a further property is necessary:
namely that the system evolves according to \equ(1.1) towards a {\it
stationary state}. This means that for all initial particles
configurations $(\dot{\V X},\V X)$, {\it except possibly for a set of
zero phase space volume}, {\it any} smooth function $f(\dot{\V X},\V
X)$ evolves in time so that, if $S_t(\dot{\V X},\V X)$ denotes the
configuration into which the initial data evolve in time $t$ according
to \equ(1.1), then the limit

$$\lim_{T\to\io} \fra1T \ig_0^T f(S_t(\dot{\V X},\V X))\,dt
=\ig f(z)\m(dz)\Eq(1.4)$$ 
exists and is independent of $(\dot{\V X},\V X)$.  The probability
distribution $\m$ is then called the {\it SRB distribution} for the
system. The maps $S_t$ will have the group property $S_t\cdot
S_{t'}=S_{t+t'}$ and the SRB distribution $\m$ will be invariant under
time evolution.

It is important to stress that the requirement that the exceptional
configurations form just a set of zero phase volume (rather than a set
of zero probability with respect to another distribution, singular
with respect to the phase volume) is a strong assumption and {\it it
should be considered an axiom of the theory}: it corresponds to the
assumption that the initial configuration is prepared as a typical
configuration of an equilibrium state, which by the classical
equidistribution axiom of equilibrium statistical mechanics is a
typical configuration with respect to the phase volume .

For this reason the SRB distribution is said to describe a stationary
state of the system. The SRB distribution depends on the parameters on
which the forces acting on the system depend, \eg $|C_0|$ (volume), 
$\BF$ (strength of the forcings), $\{\b_a^{-1}\}$ (temperatures)
\etc. The collection of SRB distributions obtained by letting the
parameters vary defines a {\it nonequilibrium ensemble}.

In the stochastic case the distribution $\m$ is required to be
invariant in the sense that it can be regarded as a marginal
distribution of an invariant distribution for the larger
(deterministic) system formed by the thermostats and the system
itself.

\*
\0{\it References}: [EM90], [Ru97a], [EPR99].
\*
%\pagina

\Section(2, Nonequilibrium thermodynamics)
\*

The key problem of nonequilibrium statistical mechanics is to derive a
macroscopic ``nonequilibrium thermodynamics'' in a way similar to the
derivation of equilibrium thermodynamics from equilibrium statistical
mechanics.

The first difficulty is that nonequilibrium thermodynamics is not well
understood. For instance there is no (agreed upon) definition of {\it
entropy}, while it should be kept in mind that the effort to find the
microscopic interpretation of equilibrium entropy, as defined by
Clausius, was a driving factor in the foundations of equilibrium
statistical mechanics.

The importance of entropy in classical equilibrium thermodynamics
rests on the implication of universal, parameter free,  relations which
follow from its existence ({\it e.g.} $\dpr_V \fra1T\=\dpr_U\fra{p}T$ if
$U$ is the internal energy, $T$ the absolute temperature and $p$ the
pressure of a simple homogeneous material).

Are there universal relations among averages of observables with
respect to SRB distributions?

The question has to be posed for systems ``really'' out of
equilibrium, \ie for $\BF\ne\V0$ (see \equ(1.1)): in fact there is a
well developed theory of the derivatives with respect to $\BF$ of
averages of observables evaluated at $\BF=\V 0$. The latter theory is
often called, and here we shall do so as well, {\it classical
nonequilibrium thermodynamics} or {\it near equilibrium
thermodynamics} and it has been quite successfully developed on the
basis of the notions of equilibrium thermodynamics, paying particular
attention on the macroscopic evolution of systems described by
macroscopic continuum equations of motion.

``Stationary nonequilibrium statistical mechanics'' will indicate a
theory of the relations between averages of observables with respect
to SRB distributions. Systems so large that their volume elements can
be regarded as being in locally stationary nonequilibrium states could
also be considered. This would extend the familiar ``local equilibrium
states'' of classical nonequilibrium thermodynamics: however they are
not considered here.  This means that we shall not attempt at finding
the macroscopic equations regulating the time evolution of continua
locally in nonequilibrium stationary states but we shall only try to
determine the properties of their ``volume elements'' assuming that
the time scale for the evolution of large assemblies of volume
elements is slow compared to the time scales necessary to reach local
stationarity.
\*
{\it References}: [DGM84], [Le93], [Ru97a], [Ru99a], [Ga98], [GL03], [Ga04].
\*

\Section(3, Chaotic hypothesis)
\*

In equilibrium statistical mechanics the {\it ergodic hypothesis}
plays an important conceptual role as it implies that the motions of
ergodic systems have a SRB statistics and that the latter coincides
with the Liouville distribution on the energy surface.

An analogous role has been proposed for the {\it chaotic hypothesis}:
which states that the {\it motion of a chaotic system, developing on
its attracting set, can be regarded as an Anosov system}. This means
that the attracting sets of chaotic systems, physically defined as systems
with at least one positive Lyapunov exponent, can be regarded as
smooth surfaces on which motion is highly unstable:
\*

\0(i) around every point a curvilinear coordinate system can be
established which has three planes, varying continuously with $x$,
which are {\it covariant} (\ie are coordinate planes at a point $x$
which are mapped, by the evolution $S_t$, into the corresponding
coordinate planes around $S_t x$) and
\\
(ii) the planes are of three types, {\it stable, unstable and
marginal}, with respective positive dimensions $d_s,d_u$ and $1$:
lengths on the stable surface and on the unstable surface of any point
contract at exponential rate as time proceeds towards the future or
towards the past. The length along the marginal direction neither
contracts nor expands (\ie it varies around the initial value staying
bounded away from $0$ and $\io$): its tangent vector is parallel to
the flow. In cases in which time evolution is discrete, and determined
by a map $S$, the marginal direction is missing.
\\
(iii) the contraction over a time $t$, positive for lines on the
stable plane and negative for those on the unstable plane, is
exponential, \ie lengths are contracted by a factor uniformly 
bounded by $C e^{-\k |t|}$ with $C,\k>0$.
\\
(iv) there is a dense trajectory.

\kern2pt
It has to be stressed that the chaotic hypothesis concerns physical
systems: mathematically {\it it is very easy to find dynamical systems
for which it does not hold}. As it is easy (actually even easier) to
find systems in which the ergodic hypothesis does not hold (\eg
harmonic lattices or black body radiation). However, if suitably
interpreted, the ergodic hypothesis leads even for these systems to
physically correct results (the specific heats at high temperature,
the Raileigh-Jeans distribution at low frequencies).  Moreover the
failures of the ergodic hypothesis in physically important systems
have led to new scientific paradigms (like quantum mechanics from the
specific heats at low temperature and Planck's law).

Since physical systems are almost always not Anosov systems it is very
likely that probing motions in extreme regimes will make visible the
features that distinguish Anosov systems from non Anosov systems: much
as it happens with the ergodic hypothesis.

The interest of the hypothesis is to provide a framework in which
properties like the existence of an SRB distribution is \ap
guaranteed: the role of Anosov systems in chaotic dynamics is similar
to the role of harmonic oscillators in the theory of regular motions.
They are the paradigm of chaotic systems as the harmonic oscillators
are the paradigm of order. Of course the hypothesis is only a
beginning and one has to learn how to extract information from it, as
it was the case with the use of the Liouville distribution once the
ergodic hypothesis guaranteed that it was the appropriate distribution
for the study of the statistics of motions in equilibrium situations.
\*
\0{\it References}: [Ru76], [GC95], [Ru97a], [Ga98], [GBG04].
\*
%\pagina

\Section(4, Heat, temperature and entropy production)
\*

The amount of heat $\dot Q$ that a system produces while in a stationary state
is naturally identified with the work that the thermostat forces
$\Bth$ perform per unit time

$$\dot Q=\sum_i \Bth_i\cdot\dot{\V x}_i\Eq(4.1)$$
A system may be in contact with several reservoirs:
in models this will be reflected by a decomposition

$$\Bth=\sum_{a=1}^m \Bth^{(a)}(\dot{\V X},\V X)\Eq(4.2)$$
where $\Bth^{(a)}$ is the force due to the $a$-th thermostat and depends
on the coordinates of the particles which are in a region
$\L_a\subseteq C_0$ of a decomposition $\cup_{a=1}^m \L_a= C_0$ of the
container $C_0$ occupied by the system ($\L_a\cap\L_{a'}=\emptyset$ if
$a\ne a'$). 

From several studies based on simulations of thermostated systems of
particles arose the proposal to consider the average of the phase space 
contraction $\s^{(a)}(\dot{\V X},\V X)$ due to the $a$-th thermostat

$$\s^{(a)}(\dot{\V X},\V X)\defi \sum_j
\dpr_{\dot{\V x}_j}\cdot \Bth^{(a)}_j (\dot{\V X},\V X)\Eq(4.3)$$
and to identify it with the {\it rate of entropy creation} in the $a$-th
thermostat. 

Another key notion in thermodynamics is the {\it temperature} of a
reservoir; in the infinite deterministic thermostats case, of Sect.\sec(1),
it is defined as $(k_B
\b_a)^{-1}$ but in the finite determinstic thermostats
considered here it needs to be defined. If there are $m$ reservoirs
with which the system is in contact one sets

$$\eqalign{
\s^{(a)}_+&\defi
\media{\s^{(a)}(\dot{\V X},\V X)}\=\ig \s^{(a)}(\dot{\V X},\V X)
\,\m(d \dot{\V X}\,d\V X)\cr
\dot Q_a&\defi\sum_i \Bth^{(a)}_i\cdot\dot{\V x}_i\cr}\Eq(4.4)$$
where $\m$ is the SRB distribution describing the stationary state. It
is natural to define, if $k_B$ is Boltzmann's constant,
the {\it absolute temperature} of the $a$-th
thermostat to be

$$T_a=\fra{\media{\dot Q_a}}{k_B\s^{(a)}_+}.\Eq(4.5)$$

Although it is known that $\fra{\sum_a \media{\dot Q_a}}{\sum_a
\s^{(a)}_+}\ge0$ it is not clear that
$T_a>0$: this happens in a rather general class of models and it would
be desirable, for the interpretation that is proposed here, that it
could be considered a property to be added to the requirements that
the forces $\Bth^{a}$ be thermostats models.

An important class of thermostats for which the property $T_a>0$ holds
can be described as follows. Imagine $N$ particles in a container
$C_0$ interacting via a potential $V_0=\sum_{i<j} \f (\V q_i-\V
q_j)+\sum_j V'(\V q_j)$ (where $V'$ models external conservative
forces like obstacles, walls, gravity, $\ldots$) and, furthermore,
interacting with $M$ other systems $\Si_a$, of $N_a$ particles of mass
$m_a$, in containers $C_a$ contiguous to $C_0$. The latter will model
$M$ parts of the system in contact with thermostats at temperatures
$T_a$, $a=1,\ldots,M$.

The coordinates of the particles in the $a$-th system $\Si_a$ will be
denoted $\V x^a_j, \, j=1,\ldots,N_a$, and they will interact with
each other via a potential $V_a=\sum_{i,j}^{N_a}
\f_a(\V x^a_i-\V x^a_j)$. Furthermore there will be an interaction
between the particles of each thermostat and those of the system via
potentials $W_a=\sum_{i=1}^N\sum_{j=1}^{N_a} w_a(\V q_i-\V
x^a_j)$, $a=1,\ldots,M$.

The potentials will be assumed to be either hard core or non singular
potentials and the external $V'$ is supposed to be at least such that
it forbids existence of obvious constants of motion.

The temperature of each $\Si_a$ will be defined by the total kinetic
energy of its particles, \ie by $K_a=\sum_{j=1}^{N_a} \fra12 m_a (\dot{\V
x}^a_j)^2\defi \fra32 N_a k_B T_a$: the particles of the $a$-th
thermostat will be kept at constant temperature by further forces
$\Bth^a_j$.  The latter are defined by imposing via a Gaussian
constraint that $K_a$ is a constant of motion (see \equ(1.3) with
$\ps\= K_a$) . This means that the equations of motion are

$$\eqalign{m \,\ddot{\V q}_j&= -\dpr_{\V q_j} \big(V_0(\V Q) +\sum_{a=1}^{N_a}
W_a(\V Q,\V x^a)\big)\cr
m_a\, \ddot{\V x}^a_j&=  -\dpr_{\V x^a_j} \big( V_a(\V x^a)+W_a(\V Q,\V
x^a)\big)-
\Bth^a_j\cr}
\Eq(4.6)$$
and an application of Gauss' principle yields $\Bth_j^a=
\fra{L_a-\dot V_a} {3N_a k_B T_a}\,\,\dot{\V x}^a_j\defi \a^a \,\dot{\V
x}^a_j $ where $L_a$ is the work per unit time done by the particles
in $C_0$ on the particles of $\Si_a$ and $V_a$ is their potential
energy.

In this case the partial divergence $\s^a=3 N_a \a^a=
\fra{L_a}{k_B T_a}-
\fra{\dot V_a}{k_B T_a}$ will make \equ(4.5) identically satisfied
with $T_a>0$ because $L_a$ can be naturally interpreted as {\it heat}
$Q_a$ ceded, per unit time, by the particles in $C_0$ to the subsystem
$\Si_a$ (hence to the $a$-th thermostat because the temperature of
$\Si_a$ is constant), while the derivative of $V_a$ {\it will not
contribute to the value of} $\s^a_+$.  The phase space contraction
rate is, {\it neglecting the total derivative terms},
$$\s_{true}(\dot{\V X},\V X) = \sum_{a=1}^{N_a}
\fra{\dot Q_a}{k_B T_a}\,.\Eq(4.7)$$
where the subscript ``true'' is to remind that an additive total
derivative term distinguishes it from the complete phase space
contraction.

\0{\it Remarks:} (1) 
The above formula provides the motivation of the name ``entropy
creation rate'' attributed to the phase space contraction $\s$.  Note
that in this way the definition of entropy creation is ``reduced'' to
the equilibrium notion because what is being defined is the entropy
increase of the thermostats which have to be considered in
equilibrium. No attempt is made here to define the entropy of the
stationary state. Nor any attempt is made to define the notion of
temperature of the nonequilibrium system in $C_0$ (the $T_a$ are temperatures of
the $\Si_a$, not of the particles in $C_0$). This is an important
point as it leaves open the possibility of envisaging the notion of
``local equilibrium'' which becomes necessary in the approximation
(not considered here) in which the system is regarded as a continuum.

\0(2) In the above model another viewpoint is possible: \ie to consider the
system to consist of only the $N$ particles in $C_0$ and the $M$
systems $\Si_a$ to be thermostats. From this point of view the above
can be considered a model of a system subject to thermostats. The
Gibbs distribution characterizing the infinite thermostats of
Sect.\sec(1) becomes in this case the constraint that the kinetic
energies $K_a$ are constants, enforced by the Gaussian forces.  This
shows that the phase space contraction can be an appropriate
definition of entropy creation rate only if the system is subject to
finite deterministic thermostats. In the new viewpoint the appropriate
definition should be simply the \rhs of \equ(4.7), \ie {\it the work per
unit time done by the forces of the system on the thermostats divided
by the temperature of the thermostats}. This suggests a more general
definition of entropy creation rate, applying also to thermostats that
are often considered ``more physical'' and that needs to be further
investigated.

\*
\0{\it References:} [EM90], [GC95], [Ru96], [Ru97b], [Ga04].

\*
\Section(5, Thermodynamic fluxes and forces)
\*

Nonequilibrium stationary states depend upon external parameters
$\f_j$ like the temperatures $T_a$ of the thermostats or the size of
the force parameters $\BF=(\f_1,\ldots,\f_q)$, see\ \equ(1.1).
Nonequilibrium thermodynamics is well developed at ``low forcing'':
strictly speaking this means that it is widely believed that we
understand properties of the derivatives of the averages of
observables with respect to the external parameters {\it if evaluated
at $\f_j=0$}. Important notions are the notions of {\it thermodynamic
fluxes} $J_i$ and of {\it thermodynamic forces} $\f_i$; hence it seems
important to extend such notions to nonequilibrium systems (\ie
$\BF\ne\V0$). 

A possible extension could be to define the thermodynamic flux $J_i$
associated with a force $\f_i$ as $J_i=\media{\dpr_{\f_i} \s}_{SRB}$
where $\s(\V X,\dot{\V X}; \BF)$ is the volume contraction per unit
time. This definition seems appropriate in several concrete cases that
have been studied and it is appealing for its generality.

An interesting example is provided by the model of thermostated
system in \equ(4.6): if the container of the system is a box with
periodic boundary conditions one can imagine to add an extra constant
force $\V E$ acting on the particles in the container. Imagining the
particles to be charged by a charge $e$ and regarding such force as an
electric field the first equation in \equ(4.6) is modified by the
addition of a term $e \V E$. 
 
The constraints on the thermostats temperatures imply that
$\s$ depends also on $\V E$: in fact, if $\V J=e\sum_j \dot{\V q}_j$
is the electric current, energy balance implies $\dot U_{tot}=\V
E\cdot\V J-\sum_a (L_a-\dot V_a)$ if $U_{tot}$ is the sum of {\it all}
kinetic and potential energies. Then the phase space contraction
$\sum_a \fra{L_a-\dot V_a}{T_a}$ can be written, to first order in the
temperature variations $\d T_a$ with respect to a common value
$T_a=T$, as $ -\sum_a \fra{L_a-\dot V_a}T
\fra{\d T_a}{T}+\fra{\V
E\cdot\V J-\dot U_{tot}}{T}$ hence $\s_{true}$, see \equ(4.7), is 

$$\s_{true}=\fra{\V E\cdot \V J}{k_BT}-\sum_a
\fra{\dot Q_a}{k_B T}\fra{\d T_a}{T}
%- \fra{\dot U_{tot}}{k_B T}
\Eq(5.1)$$

The definition and extension of the conjugacy between thermodynamic
forces and fluxes is compatible with the key results of classical
nonequilibrium thermodynamics, at least as far as
Onsager reciprocity and Green-Kubo's formulae are concerned. It can be
checked that if the equilibrium system is {\it reversible}, \ie if
there is an isometry $I$ on phase space which anticommutes with the
evolution ($I S_t=S_{-t} I$ in the case of continuous time dynamics
$t\to S_t$ or $IS=S^{-1}I$ in the case of discrete time dynamics $S$)
then, shortening $(\dot{\V X},\V X)$ into $x$,

$$\eqalign{
L_{ij}\defi&\dpr_{\F_i} J_j|_{\BF=\V0}=
\dpr_{\F_i} \media{\dpr_{\F_j} \s(x;\BF)}_{SRB}|_{\BF=\V0}
=\dpr_{\F_j} J_i|_{\BF=\V0}=\cr
&=L_{ji}=\fra12\ig_{-\io}^\io
\media{\dpr_{\F_j} \s(S_t x; \BF)\,\dpr_{\F_i} \s(x;
\BF)}_{SRB}\Big|_{\BF=\V0}\, dt\cr}
\Eq(5.2)$$
The $\s( x; \BF)$ plays the role of ``Lagrangian'' generating the
duality between forces and fluxes. The extension of the duality just
considered might be of interest in situations in which $\BF\ne\V0$.
\*
\0{\it References:} [DGM84],[Ga96],[GR97].
\*

\Section(6, Fluctuations)
\*

As in equilibrium, large statistical fluctuations of observables are of great
interest and already there is, at the moment, a rather large set of
experiments dedicated to the analysis of large fluctuations in
stationary states out of equilibrium.

If one defines the dimensionless phase space contraction

$$p(x)=\fra1\t \ig_0^\t \fra{\s(S_t x) }{\s_+}\, dt\Eq(6.1)$$
(see also \equ(4.7)) then there exists $p^*\ge1$ such that the
probability $P_\t$ of the event $p\in [a,b]$ with $[a,b]\subset
(-p^*,p^*)$ has the form

$$P_\t(p\in[a,b])\,=\,const\, e^{\t \max_{p\in [a,b]} \z(p)
+O(1)}\Eq(6.2)$$
with $\z(p)$ analytic in $(-p^*,p^*)$. The function $\z(p)$ can be 
conveniently normalized to have value $0$ at $p=1$ (\ie at the average
value of $p$).

Then, {\it in Anosov systems which are reversible and dissipative}
(see Sect.\sec(5)) a general symmetry property, called the {\it
fluctuation theorem} and reflecting the reversibility symmetry, yields
the {\it parameterless} relation

$$\z(-p)=\z(p)-p\s_+ \qquad p\in(-p^*,p^*)\Eq(6.3)$$
This relation is interesting because it has no free parameters, in
other words it is {\it universal} for reversible dissipative Anosov
systems. In connection with the duality fluxes-forces in Sec.5, it can
be checked to reduce to the Green--Kubo formula and to Onsager
reciprocity, see \equ(5.2), in the case in which the evolution depends
on several fields $\BF$ and $\BF\to\V0$ (of course the relation
becomes trivial as $\BF\to\V0$ because $\s_+\to0$ and to obtain the
result one has first to divide both sides by suitable powers of the fields
$\BF$).

A more informal (but imprecise)  way of writing \equ(6.2),\equ(6.3) is

$$\fra{P_\t(p)}{P_\t(-p)}=e^{\t p \s_++ O(1)}, \qquad \ {\rm for\ all} \
p\in(-p^*,p^*)
\Eq(6.4)$$
where $P_\t(p)$ is the probability density of $p$. An interesting
consequence
of \equ(6.4) is $\media{e^{-\t\,p\,\s_+}}_{SRB}=1$ in the sense that 
$\fra1\t \log \media{e^{-\t\,p\,\s_+}}_{SRB}\tende{\t\to\io}0$.

Occasionally systems with singularities have to be considered: in such
cases the relation \equ(6.3) may change in the sense that the function
$\z(p)$ may be not analytic: in such cases one expects that the
relation holds in the largest analyticity interval symmetric around
the origin. In various cases considered in the literature such
interval appears to contain the interval $(-1,1)$ and sometimes this
can be proved rigorously. For instance in simple, although admittedly
special, examples of systems close to equilibrium.

It is important to remark that the above fluctuation relation is the
first representative of remarkable consequences of the reversibility
and chaotic hypotheses. For instance given $F_1,\ldots,F_n$ {\it
arbitrary} observables which are (say) odd under time reversal $I$ (\ie
$F(I x)=-F(x)$) and given $n$ functions $t\in[-\fra\t2,\fra\t2]\to
\f_j(t)$, $j=1,\ldots,n$ one can ask which is the probability that
$F_j(S_tx)$ ``closely follows'' the {\it pattern} $\f_j(t)$ and at the
same time $\fra1\t\ig_0^\t
\fra{\s(S_\th x)}{\s_+}\,d\th$ has value $p$. Then calling
$P_\t(F_1\sim\f_1,\ldots,F_n\sim\f_n,p)$ the probability of this event,
which we write in the imprecise form corresponding to \equ(6.4) for
simplicity, and defining $I\f_j(t)\defi-\f_j(-t)$ it is

$$\fra{P_\t(F_1\sim\f_1,\ldots,F_n\sim\f_n,p)}
{P_\t(F_1\sim I\f_1,\ldots,F_n\sim I\f_n,-p)}
=e^{\t \s_+ p}, \qquad p\in (-p^*,p^*)\Eq(6.5)$$
which is remarkable because it is parameterless and at the same time
surprisingly independent of the choice of the observables $F_j$.  The
relation \equ(6.5) has far reaching consequences: for instance if
$n=1$ and $F_1=\dpr_{\F_i}\s(x;\BF)$ the relation \equ(6.5) has been
used to derive the mentioned Onsager reciprocity and Green--Kubo's
formulae at $\BF=\V0$. 

Eq. \equ(6.5) can be read as follows: the probability that the
observables $F_j$ follow given {evolution patterns} $\f_j$ conditioned
to entropy creation rate $p\s_+$ is {\it the same} that they follow
the time reversed patterns if conditioned to entropy creation rate
$-p\s_+$. In other words to change the sign of time it is {\it just}
sufficient to reverse the sign of entropy creation rate, no ``extra
effort'' is needed.
\*
\0{\it References:} [Si72],[Si94],[ECM93],%
[GC85],[Ga96],[GR97],[Ga99],[GBG04],[BGGZ05].
\*

\Section(7, Fractal attractors. Pairing. Time reversal.)
\*

Attracting sets (\ie sets which are the closure of attractors) are
fractal in most dissipative systems. However the chaotic hypothesis
assumes that fractality can be neglected. Aside from the very
interesting cases in which systems are close to equilibrium, in which
the closure of an attractor is the whole phase space (under the
chaotic hypothesis, \ie if the system is Anosov) there are, however,
serious problems in preserving validity of the fluctuation theorem.

The reason is very simple: if the attractor closure is smaller than
phase space then it is to be expected that time reversal will change
the attractor into a repeller disjoint from it.  Thus even if the
chaotic hypothesis is assumed, so that the attracting set $\AA$ can be
considered a smooth surface, the motion on the attractor will not be
time reversal symmetric (as its time reversal image will develop on
the repeller): one can say that an attracting set with dimension lower
than that of phase space in a time reversible system corresponds to a
{\it spontaneous breakdown of time reversal symmetry}.

It has been noted however that there are classes of systems, forming a
large set in the space of evolutions depending on a parameter $\F$, in
which geometric reasons imply that if beyond a critical value $\F_c$
the attracting set becomes smaller than phase space then a map $I_P$
is generated mapping the attractor $\AA$ into the repeller $\RR$, and
{\it viceversa}, such that $I_P^2$ is the identity on $\AA\cup\RR$ and $I_P$
{\it commutes with the evolution}: therefore the composition $I\cdot
I_P$ is a time reversal symmetry (\ie it anticommutes with evolution)
for the motions on the attracting set $\AA$ (as well as on the
repeller $\RR$). 

In other words the time reversal symmetry in such systems ``cannot be
broken'': if spontaneous breakdown occurs (\ie $\AA$ is not mapped into
itself under time reversal $I$) a new symmetry $I_P$ is spawned and
$I\cdot I_P$ is a new time reversal symmetry (an analogy with the
spontaneous violation of time reversal in quantum theory, where time
reversal $T$ is violated but $TCP$ is still a symmetry: so $T$ plays
the role of $I$ and $CP$ that of $I_P$).

Thus a fluctuation relation will hold for the phase space contraction
of the motions taking place on the attracting set for the class of
systems with the geometric property mentioned above (technically the
latter is called {\it axiom C} property). This is interesting but it
still is quite far from being checkable even in numerical experiments.

There are nevertheless systems in which also a {\it pairing} property
holds: this means that, considering the case of discrete time maps
$S$, the Jacobian matrix $\dpr_x S(x)$ has $2N$ eigenvalues that can
be labeled, in decreasing order, $\l_N(x),\ldots,\l_{\fra12 N}(x),$
$\ldots,\l_1(x)$ with the remarkable property that
$\fra12(\l_{N-j}(x)+\l_j(x))\defi\a(x)$ is $j$--independent. In such
systems a relation can be established between phase space
contractions in the full phase space and on the surface of the
attracting set: the fluctuation theorem for the motion on the
attracting set can therefore be related to the properties of the
fluctuations of the total phase space contraction measured on the
attracting set (which includes the contraction transversal to the
attracting set) and if $2M$ is the attracting set dimension and $2N$
is the total dimension of phase space it is, in the analyticity
interval $(-p^*,p^*)$ of the function $\z(p)$,

$$\z(-p)=\z(p)-p \fra{M}N \s_+\Eq(7.1)$$
which is an interesting relation. It is however very difficult to
test in mechanical systems because in such systems it seems very
difficult to make the field so high to see an attracting set thinner
than the whole phase space and {\it still} observe large
fluctuations.
\*

\0{\it References:} [DM96], [Ga99].
\*

\*
\Section(8, Nonequilibrium ensembles and their equivalence)
\*

Given a chaotic system the collection of the SRB distributions
associated with the various control parameters (volume, density,
external forces, ...) forms an ``ensemble'' describing the possible
stationary states of the system and their statistical properties.

As in equilibrium one can imagine that the system can be described
equivalently in several ways at least when the system is large (``in
the thermodynamic'' or ``macroscopic limit''). In
nonequilibrium equivalence can be quite different and more structured
than in equilibrium because one can imagine to change not only the
control parameters but also the thermostatting mechanism.

It is intuitive that a system may behave in the same way under the
influence of different thermostats: the important phenomenon being the
extraction of heat and not the way in which it is extracted from the
system. Therefore one should ask when two systems are ``physically
equivalent'', \ie when the SRB distributions associated with them give
the same statistical properties for the same observables, at least for
the {\it very few} observables which are macroscopically relevant. The
latter may be a few more than the usual ones in equilibrium
(temperature, pressure, density, ...) and include currents,
conducibilities, viscosities, ... but they will always be very few
compared to the (infinite) number of functions on phase space.

As an example consider a system of $N$ interacting particles (say hard
spheres) of mass $m$ moving in a periodic box $C_0$ of side $L$
containing a regular array of spherical scatterers (a basic model for
electrons in a crystal) which reflect particles, elastically and are
arranged so that no straight line exixts in $C_0$ which avoids the
obstacles (to eliminate obvious constants of motion). An external
field $E \V u$ acts also along the $\V u$-direction: hence the
equations of motion are

$$ m\ddot{\V x}_i=\V f_i+ E\V u-\Bth_i\Eq(8.1)$$
where $\V f_i$ are the interparticle and scatterers-particles forces
and $\Bth_i$ are the thermostating forces. The following thermostats
models have been considered

\0(1) $\Bth_i= \n\,\dot {\V x}_i$  ({\it viscosity thermostat})

\0(2) immediately after elastic collision with an obstacle the
velocity is rescaled to a prefixed value $\sqrt{3 k_B Tm^{-1}}$ for some $T$
({\it Drude's thermostat})

\0(3) $\Bth_i=\fra{\V E\cdot \sum \dot{\V x}_i}{\sum_i\dot{\V x}_i^2}$
({\it Gauss' thermostat})

The first two are not reversible. At least not manifestly such,
because the natural time reversal, \ie change of velocity sign, is not
a symmetry (there might be however more hidden, hitherto unknown, symmetries
which anticommute with time evolution). The third is reversible and
time reversal is just the change of the velocity sign. The third
thermostat model generates a time evolution in which the total kinetic
energy $K$ is constant.

Let $\m'_\n,\m''_T,\m'''_K$ be the SRB distributions for the system in
a container $C_0$ with volume $|C_0|=L^3$ and density
$\r=\fra{N}{L^3}$ fixed.  Imagine to tune the values of the control
parameters $\n,T,K$ in such a way that $\media{\rm kinetic\
energy}_\m=\EE$, {\it with the same} $\EE$ for
$\m=\m'_\n,\m''_T,\m'''_K$ and consider a {\it local} observable
$F(\dot{\V X},\V X)>0$ depending only on the coordinates of the
particles located in a region $\L\subset C_0$. Then a reasonable
conjecture is that

$$\lim_{L\to\io\atop \fra{N}{L^3}=\r}
\fra{\media{F}_{\m'_\n}}{\media{F}_{\m''_T}}=
\lim_{L\to\io\atop \fra{N}{L^3}=\r}
\fra{\media{F}_{\m'_\n}}{\media{F}_{\m'''_T}}=1\Eq(8.2)$$
if the limits are taken at fixed $F$ (hence at fixed $\L$ while
$L\to\to\io$).  The conjecture is an open problem: it illustrates,
however, the a kind of questions arising in nonequilibrium statistical
mechanics.

\*
\0{\it References:} [ES93],[Ga99],[Ru99b].

\*
\Section(9, Outlook)
\*

\0The subject is (clearly) at a very early stage of development. 
\*

(1) The theory can be extended to stochastic thermostats quite
    satisfactorily, at least as far as the fluctuation theorem is
    concerned.

(2) Remarkable works have appeared on the theory of systems which are
    purely Hamiltonian and (therefore) with thermostats that are
    infinite: unfortunately the infinite thermostats can be treated,
    so far, only if the systems are ``free'' at infinity (either free
    gases or harmonic lattices).

(3) The notion of entropy turns out to be extremely difficult to
    extend to stationary states and there are even doubts that it could
    be actually extended. Conceptually this is certainly a major open
    problem.

(4) The statistical properties of stationary states out of equilibrium
    are still quite mysterious and surprising: recently appeared exactly
    solvable models as well as attempts at unveiling the deep reasons
    for their solubility and at deriving from them general guiding
    principles. 

(5) Numerical simulations have given a strong impulse to the subject,
    in fact one can even say that they created it: introducing the
    notion of thermostat and providing the first reliable results on
    the properties of systems out of equilibrium. Simulations continue
    to be an essential part of the effort of research on the field.

(6) Approach to stationarity leads to many important questions: is
    there a Lyapunov function measuring the distance between an
    evolving state and the stationary state towards which it evolves?
    In other words can one define an analogous of Boltzmann's
    $H$-function? About this question there have been proposals 
    and the answer seems affirmative, but it does not seem
    that it is possible to find a universal, system independent,
    such function (search for it is related to the problem of defining an
    entropy function for stationary states: its existence is at least
    controversial, see Sec.2,3). 

(7) How irreversible is a given {\it irreversible process} in which
    the initial state $\m_0$ is a stationary state and at time $t=0$
    and the external parameters $\BF_{0}$ start changing into functions
    $\BF(t)$ of $t$ and tend to a limit $\BF_{\io}$ as $t\to\io$? In
    this case the stationary distribution $\m_0$ starts changing and
    becomes a function $\m_t$ of $t$ which {\it is not stationary} but
    approaches another stationary distribution $\m_\io$ as
    $t\to\io$. The process is, in general, irreversible and the
    question is how to measure its {\it degree of irreversibility}. A
    natural quantity $\II$ associated with the evolution from an
    initial stationary state to a final stationary state through a
    change in the control parameters can be defined as
    follows. Consider the distribution $\m_t$ into which $\m_0$
    evolves in time $t$, and consider also the SRB distribution
    $\m_{\BF(t)}$ corresponding to the control parameters ``frozen''
    at the value at time $t$, \ie $\BF(t)$. Let the phase space
    contraction, when the forces are ``frozen'' at the
    value $\BF(t)$, be $\s_t(x)=\s(x;\BF(t))$. In general $\m_t\ne
    \m_{\BF(t)}$. Then

   $$\II(\{\BF(t)\},\m_{0},\m_{\io})\defi \ig_0^\io
   (\m_t(\s_t)-\m_{\BF(t)}(\s_t))^2\,dt\Eq(9.1)$$
can be called the {\it degree of irreversibility of the process}: it has the
property that in the limit of infinitely slow evolution of $\BF(t)$,
\eg if $\BF(t)=\BF_0+(1- e^{-\g \k t})\BD $ (a {\it quasi static
evolution} on time scale $\g^{-1}\k^{-1}$ from $\BF_0$ to
$\BF_\io=\BF_0+\BD$), the irreversibility degree
$\II_\g\tende{\g\to0}0$ if (as in the case of Anosov evolutions, hence
under the chaotic hypothesis) the approach to a stationary state is
exponentially fast at fixed external forces $\BF$.
\*

The entire subject is dominated by the initial insights of Onsager on
    classical nonequilibrium thermodynamics: which concern the
    properties of the infinitesimal deviations from equilibrium (\ie
    averages of observables differentiated with respect to the control
    parameters $\BF$ and evaluated at $\BF=\V0$). The present efforts are
    devoted to studying properties at $\BF\ne\V0$. In this direction the
    classical theory provides certainly firm constraints (like Onsager
    reciprocity or Green-Kubo relations or fluctuation dissipation
    theorem) but  at a technical level it gives little help to enter
    the {\it terra incognita} of nonequilibrium thermodynamics of
    stationary states.

\*
\halign{#&#\hfill\cr
\0{\it References:}&\ [Ku98],[LS99],[Ma99]; [EPR99],[BLR00],[EY05];
[DLS01];[BDGJL01]; \cr
&
[EM90],[ECM33]; [GL03],[Ga04],[BGGZ05].\cr}
\*

\0{\it Bibliography}
\*

\def\*{\vskip1mm}

\0[BDGJL01] L. Bertini, A. De Sole, D. Gabrielli, G. Jona, C. Landim:
{\it Fluctuations in Stationary Nonequilibrium States of Irreversible
Processes}, Physical Review Letters, {\bf??}, , 2001.87 (2001) 040601

\*\0[BLR00] F. Bonetto, J.L. Lebowitz, L. Rey-Bellet: {\it Fourier's
law: a challenge to theorists}, in ``Mathematical Physics 2000'', p. 128--150,
editor R. Streater, Imperial College Press, 2000.

\*\0[BGGZ05] F. Bonetto, G. Gallavotti, A. Giuliani, F. Zamponi:
{\it Chaotic Hypothesis, Fluctuation Theorem and Singularities},
mp$\_$arc 05-257, cond-mat/0507672

\*\0[DM96] C. Dettman, G.P. Morriss: {\it Proof of conjugate pairing
for an isokinetic thermostat}, Physical Review {\bf 53 E}, 5545--5549,
1996.

\*\0[DGM84] S. de Groot, P., Mazur: {\it Non equilibrium
thermodynamics}, Dover, 1984, (reprint).

\*\0[DLS01] B. Derrida, J.L. Lebowitz, E. Speer: {\it Free Energy
Functional for Nonequilibrium Systems: An Exactly Solvable Case},
Physical Review Letters, 2001.

\*\0[ECM93]  D.J. Evans, E.G.D. Cohen, G., Morriss:
{\it Probability of second law violations in shearing steady flows},
Physical Review Letters, {\bf 70}, 2401--2404, 1993.

\*\0[EM90] D.J. Evans and G.P. Morriss,
{\it Statistical Mechanics of Nonequilibrium Fluids}, Academic Press,
1990, New-York.

\*\0[EPR99] J.P. Eckmann, C.A. Pillet, L. Rey Bellet:
{\it Non-Equilibrium Statistical Mechanics of Anharmonic Chains Coupled to
Two Heat Baths at Different Temperatures}, 
Communications in Mathematical Physics, {\bf 201}, 657 - 697, 1999.

\*\0[ES93] D.J. Evans, S. Sarman: {\it Equivalence of
thermostatted nonlinear responses}, Physical Review, {\bf E 48},
65--70, 1993.

\*\0[EY05] J.P. Eckmann, L.S. Young: {\it Nonequilibrium energy
profiles for a class of 1D models}, Communications in Mathematical
Physics, {\bf??}, ????, 2005.

\*\0[Ga96] G. Gallavotti: {\it Chaotic hypothesis: Onsager reciprocity
   and fluctuation--dissipation theorem}, Journal of Statistical
   Physics, {\bf 84}, 899--926, 1996.

\*\0[Ga98] G. Gallavotti, 
{\it Chaotic dynamics, fluctuations, non-equilibrium ensembles},
  Chaos, {\bf 8}, 384--392, 1998.

\*\0[Ga99] G. Gallavotti: {\it Statistical Mechanics}, Springer Verlag,
Berlin, 1999.

\*\0[Ga04] G. Gallavotti, 
{\it Entropy production in nonequilibrium stationary states: a
point of view}, Chaos, {\bf 14}, 680--690, 2004.

\*\0[GBG04] G. Gallavotti, F. Bonetto, G. Gentile:  {\it
Aspects of the ergodic, qualitative and statistical theory of
motion}, Springer--Verlag, p.1--434, 2004.

\*\0[GC95] G. Gallavotti, E.G.D. Cohen: {\it Dynamical
ensembles in non-equilibrium statistical mechanics}, Physical Review
Letters, {\bf74}, 2694--2697, 1995.

\*\0[GR97] G. Gallavotti, D. Ruelle: {\it SRB states and 
nonequilibrium statistical mechanics close to equilibrium},
   Communications in Mathematical Physics, {\bf190} (1997), 279--285.

\*\0[GL03] S. Goldstein, J. Lebowitz,
{\it On the ({B}oltzmann) entropy of Nonequilibrium systems}, Physica
D, {\bf 193}, 53--66, 2004.

\*\0[Ku98] J. Kurchan, Fluctuation theorem for stochastic dynamics,
   Journal of Physics A, 31, 3719--3729, 1998.

\*\0[Le93] J.L. Lebowitz: {\it Boltzmann's entropy
and time's arrow}, Physics Today, 32--38, 1993 (september). 

\*\0[LS99] J.L. Lebowitz, H. Spohn: {\it The Gallavotti-Cohen Fluctuation
Theorem for Stochastic Dynamics}, Journal of Statistical
   Physics, {\bf 95}, 333--365, 1999.

\*\0[Ma99] C. Maes: {\it The Fluctuation Theorem as a {G}ibbs Property},
   Journal of Statistical Physics, {\bf 95}, 367--392, 1999.

\*\0[Ru76] D. Ruelle: {\it A measure associated with axiom $A$
attractors}, American Journal of Mathematics {\bf98}, 619--654, 1976.

\*\0[Ru96] Ruelle, D.: {\it Positivity of entropy production in
nonequilibrium statistical mechanics}, Journal of Statistical Physics,
{\bf 85}, 1--25, 1996.

\*\0[Ru97a] Ruelle, D.: {\it Smooth dynamics and new theoretical
ideas in non-equilibrium statistical mechanics}, Journal of
Statistical Physics, {\bf 95}, 393--468, 1999.

\*\0[Ru97b] D. Ruelle: {\it Entropy production in
nonequilibrium statistical mechanics}, Communications in Mathematical
Physics, {\bf189}, 365--371, 1997.

\*\0[Ru99a] Ruelle, D.: {\it A remark on the equivalence of isokinetic
and isoenergetic thermostats in the thermodynamic limit}, Journal of
Statistical Physics, {\bf100}, 757--763, 2000.

\*\0[Ru99b] D.  Ruelle: {\it A remark on the equivalence of isokinetic
and isoenergetic thermostats in the thermodynamic limit}, Journal of
Statistical Physics, {\bf100}, 757--763, 2000.

\*\0[Si72] Ya.G. Sinai: {\it Gibbs measures in ergodic theory},
Russian Mathematical Surveys {\bf 166} (1972), 21--69.

\*\0[Si94]  Ya.G. Sinai: {\it Topics in ergodic theory},
Princeton Mathematical Series, Vol. 44, Princeton University Press,
Princeton, 1994.

\end